# Understanding Band Gaps of Solids in Generalized Kohn-Sham Theory

John P. Perdew[1,2*], Weitao Yang[3], Kieron Burke[4], Zenghui Yang[1+], Eberhard K.U. Gross[5], Matthias Scheffler[6-7], Gustavo E. Scuseria[8-9], Thomas M. Henderson[8-9], Igor Ying Zhang[6], Adrienn Ruzsinszky[1], Haowei Peng[1], Jianwei Sun[10], Egor Trushin[11], and Andreas Görling[11]

[1]Depts. of Physics[1] and Chemistry[2], Temple University, Philadelphia, PA, USA 19122   [3]Dept. of Chemistry, Duke University, Durham, NC, USA 27708   [4]Deps. of Chtemistry and of Physics, University of California, Irvine, CA, USA 94720   [5]Max-Planck Institut für Mikrostrukturphysik, Weinberg 2, D-06120 Halle, Germany   [6]Fritz-Haber-Institut der Max-Planck-Gesellschaft, Faradayweg 4-6, D-14195 Berlin, Germany   [7]Dept. of Chemistry and Biochemistry, and Materials Department, University of California, Santa Barbara, Santa Barbara, CA 93106   [8]Dept. of Chemistry, Rice University, Houston, TX, USA, 77251   [9]Dept. of Physics and Astronomy, Rice University, Houston, TX, USA, 77251   [10]Dept. of Physics, The University of Texas at El Paso, El Paso, TX, USA, 79968   [11]Dept. of Chemistry and Pharmacy, Friedrich -Alexander Universität Erlangen-Nürnberg, Egerlandstr. 3, 91058 Erlangen, Germany

*e-mail perdew@temple.edu

*Classification:* PHYSICAL SCIENCES/Applied Physical Sciences

[+]Present address: Microsystem and Terahertz Research Center, 596 Yinhe Road, Chengdu, Sichuan Province, China 610200

*Author contributions.* JPP designed the project. All authors contributed to the concepts and writing. ZY, TMH, IYZ, and ET did the numerical calculations.

*Competing financial interests.* The authors declare no competing financial interests.

***Abstract.*** The fundamental energy gap of a periodic solid distinguishes insulators from metals and characterizes low-energy single-electron excitations. But the gap in the band-structure of the exact multiplicative Kohn-Sham (KS) potential substantially underestimates the fundamental gap, a major limitation of KS density functional theory. Here we give a simple proof of a new theorem: In generalized KS theory (GKS), the band gap of an extended system *equals* the fundamental gap for the approximate functional if the GKS *potential operator* is continuous and the density change is delocalized when an electron or hole is added. Our theorem explains how GKS band gaps from meta-generalized gradient approximations (meta-GGAs) and hybrid functionals can be more realistic than those from GGAs or even from the exact KS potential. The theorem also follows from earlier work. The band edges in the GKS one-electron spectrum are also related to measurable energies. A linear chain of hydrogen molecules, solid aluminum arsenide, and solid argon provide numerical illustrations.

***Significance.*** Semiconductors and insulators have a fundamental energy gap, and absorb light at a continuum of photon energies above this gap. They also have a band-structure of one-electron energies, and a band gap separating unoccupied from occupied one-electron states. When should these gaps be equal? It is known that they are not equal in the exact Kohn-Sham density functional theory, but are equal in commonly-used density-functional approximations such as the generalized gradient approximation (GGA). We show here that they are also equal (and improved) in higher-level approximations such as the meta-GGA or the hybrid of GGA with exact exchange, when the effective one-electron potential is not constrained to be a multiplication operator.

\body

***Band-gap problem in Kohn-Sham density functional theory.*** The most basic property of a periodic solid is its fundamental energy gap $G$, which vanishes for a metal but is positive for semiconductors and other insulators. $G$ dominates many properties. As the unbound limit of an exciton series, $G$ is an excitation energy of the neutral solid, but it is defined here as a difference of ground-state energies: If $E(M)$ is the ground-state energy for a solid with a fixed numfor I-A) ber of nuclei and $M$ electrons, and if $M = N$ for electrical neutrality, then

$$G = I(N) - A(N) = [E(N-1) - E(N)] - [E(N) - E(N+1)] \qquad (1)$$

is the difference between the first ionization energy $I(N)$ and the first electron affinity $A(N)$ of the neutral solid. While $I$ and $A$ can be measured for a macroscopic solid, they can be computed directly (as ground-state energy differences) either by starting from finite clusters and extrapolating to infinite cluster size or (for *I-A*) by starting from a finite number of primitive unit cells, with periodic boundary condition on the surface of this finite collection, and extrapolating to an infinite number. Here we shall follow both approaches, which have been discussed in a recent study (1). (The energy to remove an electron to infinite separation cannot depend upon the crystal face through which it is removed, although the energy to remove an electron to a macroscopic separation, but much smaller than the dimensions of that face, may so depend. The gap is of course a bulk property.)

Kohn-Sham density functional theory (2,3) is a formally-exact way to compute the ground-state energy and electron density of $M$ interacting electrons in a multiplicative external potential. This theory sets up a fictitious system of non-interacting electrons with the same ground-state density as the real interacting system, found by solving self-consistent one-electron Schrödinger equations. These electrons move in a *multiplicative* effective Kohn-Sham potential, the sum of the external and Hartree potentials and the derivative of the density functional for the exchange-correlation energy, $E_{xc}[n_\uparrow, n_\downarrow]$, which must be approximated. The simplest local spin density approximation (LSDA) (2) is already usefully accurate for many solids. Better still are generalized gradient approximations (GGAs) (e.g., Ref. 4), meta-GGAs (e.g., Refs. 5,6), and hybrids of GGA with exact exchange (e.g., Refs. 7,8). The additional ingredients in higher-level functionals can in principle satisfy more exact constraints, or fit data better, achieving higher accuracy. KS theory has become the most widely-used (3) method to calculate the ground-state energies, energy differences, electron densities, and equilibrium structures of molecules and solids, and, with less justification, the electronic band structures of solids. For a solid, KS theory produces a band structure, one-electron energies as functions of Bloch wavevector and band index, in which there can be a non-zero band gap,

$$g = \varepsilon^{LU} - \varepsilon^{HO},  \qquad (2)$$

the difference between the lowest-unoccupied (LU) and highest-occupied (HO) one-electron energies. We show here that, under common computational conditions for solids, $g$ equals $G$ for a given approximate functional. How close $g$ is to the experimental gap depends on how accurate the functional is for the ground-state energy difference $G$ (strongly and comparably underestimated by LSDA and GGAs, but better estimated by meta-GGAs and especially hybrids).

In principle, should the band gap $g$ equal the fundamental energy gap $G$? In the early 1980's, band structure calculations were accurate enough to show that LSDA band gaps for semiconductors were often about half the measured fundamental energy gaps. Was this a failure of the LSDA effective potential to mimic the exact KS potential, or an inability of the exact KS potential (for the neutral solid) to predict the fundamental gaps, or both?

Regarding the fundamental gap $G$ as an excitation energy, we do not expect it to equal the band gap $g$ of the exact KS potential. But thinking of it as a ground-state energy difference, we might hope that it is. Williams and von Barth (9) gave a clear argument to support this hope, based on three assumptions: (I) Janak's theorem (10,11): The one-electron energies of KS theory are derivatives of the total energy with respect to occupation number, between integer occupations, in both finite and extended systems. This is unquestionably true. (II) When an electron is added or removed from a solid, the density change is infinitesimal and periodic. This assumption, only possible for an extended system, is often true, although there may be exceptions in which added electrons or holes get stuck in localized states; see Refs. 12,13 for possible examples. (III) When an electron is added or removed, the KS potential changes only infinitesimally. This assumption seemed to follow so naturally from (II) that it was only implicit in the argument, yet assumption III is incorrect for the exact KS potential.

Other work (14-17) of the early 1980's showed that the exact Kohn-Sham potential jumps up by an additive-constant *discontinuity* when an electron is added to a neutral solid, making

$$G^{exact} = g^{exact} + xc\_discontinuity. \qquad (3)$$

The discontinuity spoils the interpretation of *g*, shifting the one-electron energies without changing the density. The KS potential is a mathematical fiction, acting on non-interacting electrons to yield the true ground-state density of the neutral solid and making the one-electron energy for the highest partly-occupied one-electron state equal to the true chemical potential $\mu = \partial E / \partial M$, which is itself discontinuous at zero temperature for an insulator when *M* crosses *N*. The xc discontinuity is absent in the LSDA and GGA approximations to the multiplicative exchange-correlation potential, for which (17)

$$G^{approx} = g^{approx}. \qquad (4)$$

In Eq. (4), *G* of Eq. (1) and *g* of Eq. (2) are evaluated with the same approximate functional. While GGA improves ground-state energies and electron densities over LSDA, both approximations yield nearly the same band gaps *g* and hence fundamental gaps *G*, excepting some special GGAs (18). It has long been known (17) that Eq. (4) is true in LSDA and GGA, and it has been suspected (e.g., Refs. 15,17) that LSDA and GGA band gaps are close to exact KS band gaps (but not to true fundamental gaps).

***Band-gap problem in generalized Kohn-Sham (GKS) theory.*** A simple, self-contained proof of our theorem will be given here. Refs. 19-22 by themselves also imply this result, as discussed in a later section.

Based mostly upon empiricism, realistic fundamental gaps for semiconductors (e.g., Refs. 23,24) have been estimated from band gaps of hybrid functionals in GKS, which is also an excellent starting point for simple quasi-particle corrections (25). A global hybrid replaces a fraction (e.g., 25% (7,26,27)) of GGA exchange with that of Hartree-Fock, and replaces the same fraction of the GGA exchange potential with that of Hartree-Fock (an integral operator, not a multiplication operator). Screened hybrids (e.g., Ref. 8) additionally screen the interelectronic Coulomb potential in the exchange term, and typically improve results for semiconductors (23).

We argue that Eq. (4) is also valid within typical approximations in GKS theory, as typically implemented, extending the argument of Williams and von Barth (9) from KS to GKS theory. Thus the improvement in the band gap that comes from using a hybrid functional reflects a corresponding improvement in the value for $G$ of Eq. (1). Our detailed argument, presented in the theoretical methods section, generalizes assumption (I) of the Williams-von Barth argument (9) from KS to GKS theory, and notes that the GKS potentials, like the LSDA and GGA and unlike the exact one, have no discontinuity under change of particle number, consistent with Refs. 19-22.

While there is a formally-exact GKS theory (28), here we view GKS as a small step out of KS theory, in which one can use nonempirical approximations to $E_{xc}$ that are constructed to satisfy the known exact constraints of KS theory. In rigorous KS density functional theory, the occupied Kohn-Sham one-electron states are demonstrably *implicit* functionals of the electron density that can be used to construct a density functional approximation, such as an *explicit* functional of the KS one-electron density matrix. For example, use the non-interacting kinetic energy density to construct a meta-GGA (e.g., Ref. 5), or use the full KS density matrix to construct the Hartree-Fock exchange energy for a global hybrid as in Ref. 7. Because the one-electron states are only implicit functionals of the density, the KS potential can be constructed only by the optimized effective potential (OEP) method (29). It is computationally easier to find the variationally optimized potential that minimizes the energy with respect to the non-interacting density matrix. The resulting GKS potential is not a multiplication operator, but is in practice continuous (does not change when one delocalized electron is added to or subtracted from a solid) and self-adjoint, for differentiable functionals of the non-interacting density matrix. It is an integral (Fock) operator (11) for hybrids, but a differential operator (30,31) for meta-GGA's, the same operator for occupied and unoccupied one-electron states.

The step outside KS to GKS barely affects the occupied one-electron states, the electron density, and the total energy, but not so the one-electron energies. This was first shown by comparing exchange-only OEP (KS) and Hartree-Fock (GKS) results for atoms (29,32), and more recently by comparing the corresponding KS and GKS implementations of meta-GGAs (exchange and correlation together) for atoms (33) and solids (31). They produce closely similar results for total energies, but the KS meta-GGA band gap is close to that of LSDA and GGA, while the GKS meta-GGA band gap is significantly larger and more realistic.

Within exchange-correlation approximations using the non-interacting density matrix, relaxing the KS demand for a multiplicative effective potential is a "practical" approximation with an unexpected benefit: It yields the interpretation of Eq. (4) for the GKS band gap of a solid, explaining how meta-GGAs and especially hybrids can improve the estimation of the fundamental energy gap of a solid: For a typical approximate functional, the GKS band gap $g$ is the ground-state energy difference $G$. Improvements in $G$ correlate at least roughly with other improvements in ground-state energy differences for integer electron numbers, relevant to atomization energies and lattice constants.

*Numerical demonstration.* Because computational effort typically scales like the cube of the number of atoms, finite three- and even two-dimensional clusters are much harder to converge to the mesoscopic length scale, so we consider as a first model a finite *one-dimensional* linear chain of realistic $H_2$ molecules. The separation between the nuclei of neighboring molecules is taken to be 1.25 times the separation between nuclei within a molecule (0.74 Å), in order to produce a gap of order 3 or 4 eV. To demonstrate our conclusions, the model does not need to be realistic, and its exact gap does not need to be known. With an even number (two) of electrons per unit cell, this system is a band insulator. We consider chains with one to 500 molecules. At large numbers $N_{mol}$ of molecules, the correction to the limit $N_{mol} \to \infty$ is (13,34) of order $1/N_{mol}$, simplifying the extrapolation. Figures 1 and 2 show that, for all tested approximate functionals, $G - g$ tends to zero as $N_{mol} \to \infty$. Table 1 shows limiting values. Within numerical accuracy, as $N_{mol} \to \infty$, $I \to -\varepsilon^{HO}$, $A \to -\varepsilon^{LU}$, and $G \to g$.

The positive ions show delocalization of the extra positive charge over the finite chain, even without periodic boundary conditions, as expected from the approximate functionals studied here. The negative ions are resonances, with negative electron affinity of the chain, captured by the finite basis set. But the resonance can evolve smoothly (35) to a bound state with positive electron affinity as the chain length grows. In contrast to the situation for atoms and molecules, the resonant one-electron states of bulk solids can be converged with respect to basis set.

Ref. 36 states without an explicit proof a major result proved here: For a hybrid functional implemented in a generalized Kohn-Sham scheme, the band gap equals the fundamental gap within the same approximation. Refs. 36 and 37 show how to calculate the fundamental gaps of real extended solids from a given functional without extrapolating from clusters of finite size (and Ref. 37 thereby finds realistic band gaps for many solids from the random phase approximation, by a method different from that of Ref. 38.) This makes it possible to demonstrate our conclusions for real three-dimensional solids using a computer code with periodic boundary conditions.

To that end, we report calculations for the semiconductor aluminum arsenide and the large-gap insulator solid argon with the PBE GGA (4) and the PBE0 hybrid (7,26,27) functionals as representatives for KS and GKS methods, via the approach of Refs. 36,37. Regular grids of $n$ x $n$ x $n$ $\bm{k}$ points containing the $\Gamma$ point are used, corresponding to a collection of $n$ x $n$ x $n$ primitive unit cells in periodic boundary conditions. For $n \to \infty$ an infinite periodic solid would be obtained, forbidding symmetry-breaking localization of the added electron or hole, which we do not expect for the solids and functionals considered here. Symmetry-breaking (forming polarons) can be captured by a related supercell approach (39). A self-consistent calculation for the neutral system yields a band gap $g$ and an energy $E(N)$. Removal of one electron from the highest occupied orbital or one-electron state (the $\bm{k}$ point at the top of the valence band), while keeping the other occupations and orbitals unchanged, yields the non-selfconsistent $E_{non-SCF}(N-1)$, while allowing orbital relaxation yields the selfconsistent $E_{SCF}(N-1)$. Contributions to the Hartree energy and Hartree potential from the zero reciprocal lattice vector are not taken into account in the charged systems, or (as usual) in the neutral ones. This long-known approach for charged systems (40) is better justified for bulk periodic solids than for other cases (41). Thus, without any code modification, a finite energy *E(N-1)* is obtained. An ionization potential *I(N)* is just the difference *E(N-1)-E(N)*, where neither energy is divided by the number of primitive unit cells. An energy *E(N+1)* is obtained analogously by adding one electron to the **k** point representing the bottom of the conduction band. From Eq. (1) the fundamental energy gaps $G_{non\text{-}SCF}$ and $G_{SCF}$, for the cases without and with orbital relaxation respectively, are calculated. Convergence with mesh size is rapid for PBE. For PBE0, convergence is accelerated by the method of Ref. 42. No physical (measurable) interpretation is intended for the gaps in Tables 2 and 3, except in the limit of large *n*.

Tables 2 and 3 show that all three gaps, *g*, $G_{non\text{-}SCF}$, and $G_{SCF}$, rapidly converge towards each other. The convergence of $G_{non\text{-}SCF}$ and $G_{SCF}$ towards each other demonstrates that orbital relaxation upon removal or addition of an electron does not play a role in infinite periodic solids, while the convergence of *g* and $G_{SCF}$ towards each other represents a numerical demonstration of the theorem of this work. Comparison with the experimental gaps in the table captions shows that, as expected, the 25% exact exchange in PBE0 can be too much for small-gap solids like AlAs, and too little for large-gap solids like Ar. As expected (15,17), the OEP or KS band gap *g* for PBE0 is closer to the PBE KS value than to the PBE0 GKS value.

***Relation to other previous work.*** The relation between GKS *frontier* orbitals and electron addition/removal energies was first shown in Refs. 19-22 for both extended and finite systems, and was demonstrated numerically for molecules in Ref. 19. Refs. 19-22 by themselves imply our main result. Ref. 19 derives the generalized Janak's theorem for a differentiable functional of the one-particle density matrix (in its Eq. 10), namely, the GKS LU/HO orbital energies are the chemical potentials for electron addition/removal for both finite and extended systems, in a way that differs from the derivation in our theoretical methods section. Refs. 19,20 show that the GKS one-electron energy gap matches the GKS derivative gap -- the discontinuity in chemical potentials for electron addition and removal (Eq. 5 of Ref. 20), which is equal to G for the exact functional and for functionals with linear behavior in *M* on either side of *N,* but generally differs from *G* for finite systems with approximate functionals (Eq. 6 of Ref. 20). Ref. 20 further shows that the $E(M)$ curves are linear over *M* on either side of *N* for approximate functionals in periodic solids, and also for non-periodic systems as $N \to \infty$ when the approximate functionals have delocalization error. But the GKS derivative gap is not equal to $G$ for non-periodic systems as $N \to \infty$ for functionals with localization error, such as hybrid functionals with high fractions of exact exchange that localize an added electron or hole (20). Combined, these statements yield our main conclusion.

For KS methods employing the optimized effective potential (OEP) method (29) to construct the exchange-correlation potential corresponding to orbital-dependent energy functionals, e.g., the exact exchange energy, the KS band gap $g$ and the fundamental energy gap $G$ are different as mentioned above. Indeed OEP potentials do not determine an additive constant because the electron number is kept fixed. If the KS band structures are adjusted by an appropriate shift of the gap, as in Ref. 36, they can be transformed into approximate quasiparticle band structures.

***Conclusions.*** The fundamental energy gap is the most basic property of a periodic solid. It cannot be found from a single Kohn-Sham band-structure calculation, even with the unattainable exact density functional. Surprisingly, high-level approximations, implemented in an efficient generalized KS scheme, yield band gaps equal to the fundamental gap for a given approximate functional. Future all-purpose non-empirical approximate functionals could predict usefully-correct gaps for most solids. The band edges (43) in the GKS one-electron spectrum, relevant to interface formation and redox catalysis, can also be interpreted as measurable energy differences, as shown by Eq. (6) and illustrated in Table 1. They can be found in principle by extrapolating the GKS one-electron energies of a slab or cluster.

Typical approximate functionals, as typically implemented, obey Eq. (4), as previously known (17) only for LSDA and GGA. For three-dimensional solids (31), there is little or no improvement in $G^{approx}$ from LSDA to GGAs, but substantially more from GGAs to fully-nonlocal functionals, where the nonlocality of the density dependence and the usefulness of the band gap $g^{approx}$ increase further from meta-GGAs to hybrids. This suggests that, in solids, the exchange-correlation effects can be more long-ranged (e.g., Ref. 23) than in atoms and small molecules.

The PBE0 and HSE06 hybrids contain 25% of exact exchange, globally or at intermediate range, chosen to yield accurate atomization energies energies for molecules and related moderate-gap systems at integer electron number. The nonlinear variation of approximate total energy with electron number between adjacent integers is a problem in finite systems, but vanishes in typical solids (1,20). PBE0 and especially HSE06 yield realistic GKS gaps for typical semiconductors. But they can over- or under-estimate gaps of other solids. For example, molecular crystals seem to need $1/\varepsilon$ of long-range exact exchange (44), where $\varepsilon$ is the dielectric constant (45).

*Appendix A: Computational methods.* The self-consistent all-electron results for the chain of hydrogen molecules reported here were found using the Gaussian code (46) with a small cc-pvDZ basis set, to speed up the hybrid calculations for the longer chains. Many results were checked with the ADF (47) (TZP basis) and FHI-aims (48) (NAO-VCC-2Z basis) codes. The effect of increasing the basis from cc-pvDZ to TZP is to increase the $N_{mol} \to \infty$ limits of $I$ and $A$ in PBE by 0.14 and 0.10 eV, respectively, and to stabilize the negative-ion resonances for some of the larger finite chains. All codes show $G - g \to \sim 0.02$ eV, which we attribute to the slow convergence of $G$ with increasing system size (Fig. 1). All extrapolations display the increase of $I$ and decrease of $A$ from LSDA to HSE06.

The AlAs and Ar calculations were carried out with the plane-wave program MCEXX (49) using norm-conserving PBE pseudopotentials generated by the code of Ref. 50 which is based on the Troullier-Martins scheme (51). The cutoffs used for the construction of the pseudopotentials are the same as those used in Ref. 37. In principle, the pseudopotential for PBE0 should be different from that for PBE, but the difference is irrelevant to our demonstration. For AlAs a lattice parameter of 5.66 Å and a plane wave cutoff of 15 a.u. were used. The corresponding values for Ar were 5.26 Å and 30 a.u.

*Appendix B: Theoretical methods.* Here, we derive the generalized Janak's theorem, and prove that the band gap and band edges of generalized Kohn-Sham theory are the appropriate ground-state energy differences, for a given approximate functional. In any constrained minimization, the Lagrange multiplier is the derivative of the minimized quantity with respect to the value of the constraint. Consider minimizing the orbital functional $E_v[\{f_j\},\{\psi_j\}]$, where $n(\vec{r}) = \sum_i f_i |\psi_i(\vec{r})|^2$ and the occupation numbers are restricted to the range $0 \leq f_i \leq 1$ with $\sum_i f_i = N$, subject to constraints $f_j \int d^3r |\psi_j(\vec{r})|^2 = f_j$ guaranteeing normalization of the occupied or partly-occupied orbitals. The Euler-Lagrange equation for this problem is $\delta\{E_v[\{f_j\},\{\psi_j\}] - \sum_i \varepsilon_i f_i \int d^3r |\psi_i(\vec{r})|^2\} = 0$, where the $\varepsilon_i$ are Lagrange multipliers. The interpretation is

$$\varepsilon_i = \partial E / \partial f_i. \tag{5}$$

This is a *generalized* Janak's theorem. The same statement and derivation (11) apply to the ungeneralized KS theory. The minimizing one-electron wavefunctions are solutions of a one-electron Schrödinger equation with an optimal variational potential operator.

Consider a GKS calculation for an extended solid with an approximate xc functional, in which the ground-state delocalizes the density of the added electron or hole over the infinite solid. The variation of the approximated $E$ is linear in $f_i$ because the relaxation effect on the optimal variational potential associated with the removal or addition of one electron is negligible. Then, by Eq. (5),

$$\begin{aligned} E(N) - E(N-1) &= \varepsilon^{HO}(N-\delta) \\ E(N+1) - E(N) &= \varepsilon^{LU}(N+\delta) \end{aligned} \tag{6}$$

where $\delta = 0^+$, and

$$I(N) - A(N) = \varepsilon^{LU}(N+\delta) - \varepsilon^{HO}(N-\delta). \tag{7}$$

Here HO and LU label the one-electron states of the $(N-\delta)$-electron system, which change only infinitesimally when $M$ increases through integer $N$. If the approximate xc potential in GKS theory has no discontinuity as the electron number crosses integer $N$, then

$$I(N) - A(N) = \varepsilon^{LU}(N) - \varepsilon^{HO}(N). \tag{8}$$

For a meta-GGA or hybrid functional, the optimum variational potential operator has been found explicitly (e.g., Eq. (7) of Ref. 30, Eq. (1.7) of Ref. 11) and is continuous. Thus within LSDA, GGA, meta-GGA, or hybrid approximations, when implemented in GKS, the band gap equals the ground-state total energy difference.

In contrast, within an *ungeneralized* KS scheme, this statement remains true in LSDA and GGA, but *not* in meta-GGA or hybrid approximations. For meta-GGA and hybrid approximations, treated in OEP, as for (15) exact KS theory,

$$I(N) - A(N) = \varepsilon_{OEP}^{LU}(N+\delta) - \varepsilon_{OEP}^{HO}(N-\delta) =$$
$$\{\varepsilon_{OEP}^{LU}(N-\delta) - \varepsilon_{OEP}^{HO}(N-\delta)\} + \{\varepsilon_{OEP}^{LU}(N+\delta) - \varepsilon_{OEP}^{LU}(N-\delta)\}, \tag{9}$$

where the first curly bracket is the OEP or KS band gap and the second is the contribution from the discontinuity (15,16) of the OEP or KS potential.

*Acknowledgments.* The work of JPP, WY, GES, AR, HP, and JS was part of the Center for the Computational Design of Functional Layered Materials, an Energy Frontier Research Center (EFRC) funded by the US Department of Energy, Office of Science, Basic Energy Sciences, under Award No. DE-SC0012575. JPP and AR acknowledge the support of the Humboldt Foundation for visits to the Fritz-Haber Institut. KB was supported by DOE grant number DE-FG02-08ER46496. ET and AG acknowledge support by the Deutsche Forschungsgemeinschaft through the Excellence Cluster "Engineering of Advanced Materials"

*References*

1. Vlček V, Eisenberg HR, Steinle-Neumann G, Kronik L, Baer R (2015) Deviations from piecewise linearity in the solid-state limit with approximate density functionals. J. Chem. Phys. 142:034107,1-10.

2. Kohn W, Sham LJ (1965) Self-consistent equations including exchange and correlation. Phys. Rev. 140:A1133-1138.

3. Van Noorden R, Maher B, Nuzzo R (2014) The Top 100 Papers. Nature 514:550-553.

4. Perdew JP, Burke K, Ernzerhof M (1996) Generalized gradient approximation made simple. Phys. Rev. Lett. 77: 38653868.

5. Sun J, Ruzsinszky A, Perdew JP (2015) Strongly constrained and appropriately normed semilocal density functional. Phys. Rev. Lett. 115: 036402,1-6.

6. Sun J, Remsing RC, Zhang Y, Sun Z, Ruzsinszky, Peng H, Yang Z, Paul A, Waghmare U, Wu X, Klein ML, Perdew JP (2016) Accurate first-principles structures and energies of diversely-bonded systems from an efficient density functional. Nature Chem. 8:831-836.

7. Perdew JP, Ernzerhof M, and Burke K (1996) Rationale for mixing exact exchange with density functional approximations. J. Chem. Phys. 105:9982-9985.

8. Heyd J, Scuseria GE, Ernzerhof M (2003, 2006) Hybrid functionals based on a screened Coulomb potential. J. Chem. Phys. 118:8207-8215; *ibid*. 124:219906-219906.

9. Williams AR, von Barth U (1983) Applications of density functional theory to atoms, molecules, and solids. in *Theory of the Inhomogeneous Electron Gas*, edited by S. Lundqvist and N.H. March, Plenum, N.Y., See section 4.1.

10. Janak JF (1978) Proof that $dE/df_i=\varepsilon_i$, in density functional theory, Phys. Rev. B 18:7165-7168.

11. Slater JC (1974) *The Self-Consistent Field in Molecules and Solids,* McGraw-Hill, N.Y.

12. Zhang IY, Jiang J, Gao B, Xu X, Luo Y (2014) RRS-PBC: a molecular approach for periodic systems. Science in China – Chemistry 57:1399-1404.

13. Vlček V, Eisenberg HR, Steinle-Neumann G, Neuhauser D, Rabani E, Baer R (2016) Spontaneous charge carrier localization in extended one-dimensional systems. Phys. Rev. Lett. 116:186401,1-6.

14. Perdew JP, Parr RG, Levy M, Balduz JL (1982) Density-functional theory for fractional particle number: Derivative discontinuities of the energy. Phys. Rev. Lett. 49:1691-1694.

15. Perdew JP, Levy M (1983) Physical content of the exact Kohn-Sham orbital energies – Band-gaps and derivative discontinuities. Phys. Rev. Lett. 51:1884-1887.

16. Sham LJ, Schlueter, M (1983) Density functional theory of the energy gap Phys. Rev. Lett. 51: 1888-1891.

17. Perdew JP (1985) Density functional theory and the band-gap problem. Int. J. Quantum Chem. S19: 497-523.

18. Armiento R, Kümmel S (2013) Orbital localization, charge transfer, and band gaps in semilocal density-functional theory. Phys. Rev. Lett. 111: 036402,1-5.

19. Cohen AJ, Mori-Sanchez P, Yang W (2008) Fractional charge perspective on the band gap in density functional theory. Phys. Rev. B 77:115123,1-6.

20. Mori-Sanchez P, Cohen AJ, and Yang W (2008) Localization and delocalization errors in density functional theory and implications for band-gap prediction. Phys. Rev. Lett. 100: 146401,1-4.

21. Mori-Sanchez P, Cohen AJ, Yang W (2009) Discontinuous nature of the exchange-correlation functional in strongly-correlated systems. Phys. Rev. Lett. 102: 066403,1-4.

22. Yang W, Cohen AJ, Mori-Sanchez P (2012) Derivative discontinuity, bandgap and lowest unoccupied molecular orbital in density functional theory. J. Chem. Phys. 136:204111,1-13.

23. Lucero MJ, Henderson TM, Scuseria GE (2012) Improved semiconductor lattice parameters and band gaps from a middle-range screened hybrid functional. J. Phys. Condens. Matt. 24: 145504,1-11.

24. Eisenberg HR, Baer R (2009) A new generalized Kohn-Sham method for fundamental band gaps in solids. Physical Chemistry Chemical Physics 11:4674-4680.

25. Fuchs F, Furthmueller J, Bechstedt F, Shishkin M, Kresse G (2007) Quasiparticle band structure based on a generalized Kohn-Sham scheme. Phys. Rev. B 76:115109,1-8.

26. Ernzerhof M, Scuseria GE (1999) Assessment of the Perdew-Burke-Ernzerhof exchange-correlation functional. J. Chem. Phys. 110:5029-5036.

27. Adamo C, Barone V (1999) Toward reliable density functional methods without adjustable parameters: The PBE0 model. J. Chem. Phys. 110:6158-6170.

28. Seidl A, Görling A, Vogl P, Majewski JA, Levy M (1996) Generalized Kohn-Sham schemes and the band gap problem. Phys. Rev. B 53:3764-3774.

29. Talman JD, Shadwick WF (1976) Optimized effective atomic central potential. Phys. Rev. A 14:36-40.

30. Neumann R, Nobes R, Handy NC (1996) Exchange functionals and potentials. Mol. Phys. 87:1-36.

31. Yang Z, Peng H, Sun J, Perdew JP (2016) More realistic band gaps from meta-generalized gradient approximations: Only in a generalized Kohn-Sham scheme. Phys. Rev. B 93:205205,1-9.

32. Görling A, Ernzerhof M (1995) Energy differences between Kohn-Sham and Hartree-Fock wavefunctions yielding the same density. Phys. Rev. A 51: 4501-4513.

33. Eich F, Hellgren M (2014) Derivative discontinuity and exchange-correlation potential of meta-GGA's in density functional theory. J. Chem. Phys. 141:224107,1-9.

34. Godby RW, White ID (1998) Density-relaxation part of the self-energy, Phys. Rev. Lett. 80:3161-3161.

35. Feuerbacher S, Sommerfeld T, Cederbaum LS (2004) Extrapolating bound state anions into the metastable domain. J. Chem. Phys. 121: 6628-6633.

36. Görling A (2015) Exchange-correlation potentials with proper discontinuities for physically meaningful Kohn-Sham eigenvalues and band structures. Phys. Rev. B 91:245120,1-10.

37. Trushin E, Betzinger M, Blügel S, Görling A (2016) Band gaps, ionization potentials, and electron affinities of periodic electron systems via the adiabatic-connection fluctuation-dissipation theorem. Phys. Rev. B 94: 075123,1-9.

38. Grüning M, Marini A, Rubio A (2006) Density functionals from many-body perturbation theory: The band gap for semiconductors and insulators. J. Chem. Phys. 124:154108,1-9.

39. Hofmann OT, Rinke P, Scheffler M, Heine G (2015) Integer vs. fractional charge at metal (/insulator) organic interfaces: Cu(/NaCl)/TCNE. ACSNano 9**:**5391-5404.

40. Bar-Yam Y, Joannopoulos JD (1984) Electronic structure and total-energy migration barriers of silicon self-interstitials. Phys. Rev. B 30: 1844-1852.

41. Richter NA, Sicolo S, Levchenko SV, Sauer J, Scheffler M (2013) Concentrations of vacancies at metal-oxide surfaces: Case study of MgO(100). Phys. Rev. Lett. 111: 045502,1-5.

42. Carrier P, Rohra S, Görling A (2007) General treatment of the singularities in Hartree-Fock and exact-exchange Kohn-Sham methods for solids. Phys. Rev. B 75: 205126.

43. Moses PG, Miao M, Yan Q, Van de Walle CG (2011) Hybrid functional investigations of band gaps and band alignments for AlN, GaN, InN, and InGaN. J. Chem. Phys. 134: 084703,1-11.

44. Refaely-Abramson S, Sharifzadeh S, Jain M, Baer R, Neaton JB, Kronik L (2013) Gap renormalization of molecular crystals from density-functional theory. Phys. Rev. B 88:081204,1-5.

45. Skone JH, Govoni M, and Galli G (2014) Self-consistent hybrid functional for condensed systems. Phys. Rev. B 89:195112,1-12.

46. Gaussian 09, Revision E.01 (2009) Frisch MJ *et al.*, Gaussian, Inc., Wallingford CT.

47. ADF2014, SCM, Theoretical Chemistry, Vrije Universiteit, Amsterdam, The Netherlands, http://www.scm.com, Baerends EJ *et al.*

48. Blum V, Gehrke R, Hanke F, Havu P, Havu V, Ren X, Reuter K, Scheffler M (2009) Ab initio molecular simulations with numeric atom-centered orbitals. Comp. Phys. Comm. 180:2175-2196.

49. Görling A, et al. (2015) Magnetization Current Exact-Exchange Code (University of Erlangen-Nürnberg, Erlangen, Germany)..

50. Engel E, Höck A, Schmid RN, Dreizler RM, Chetty N (2001) Role of the core-valence interaction for pseudopotential calculations with exact exchange. Phys. Rev. B 64: 125111,1-12.

51. Troullier N, Martins JL (1991) Efficient pseudopotentials for plane-wave calculations. Phys. Rev. B 43:8861-8869.

52. Madelung O (1996) *Semiconductors - Basic Data,* (Springer, New Yor0. 2$^{nd}$ Ed., p 94.

53. Schwentner N, et al. (1975) Photoemission from rare-gas solids: Electron energy distribution from the valence bands. Phys. Rev. Lett 34:5280531.

54. Garza AJ, Scuseria GE (2016) Predicting band gaps with hybrid functionals. J. Phys. Chem. Lett. 7:4165-4170.

**Figure Legends**

Fig. 1. The PBE GGA fundamental gap *G* and band gap *g* for a linear chain of $N_{mol}$ H$_2$ molecules. Note that *G* converges to the limit $N_{mol} \to \infty$ much more slowly than *g* does.

Fig. 2. Difference between the fundamental gap $G = I - A$ and the GKS band gap $g = \varepsilon^{LU} - \varepsilon^{HO}$ for a linear chain of $N_{mol}$ hydrogen molecule

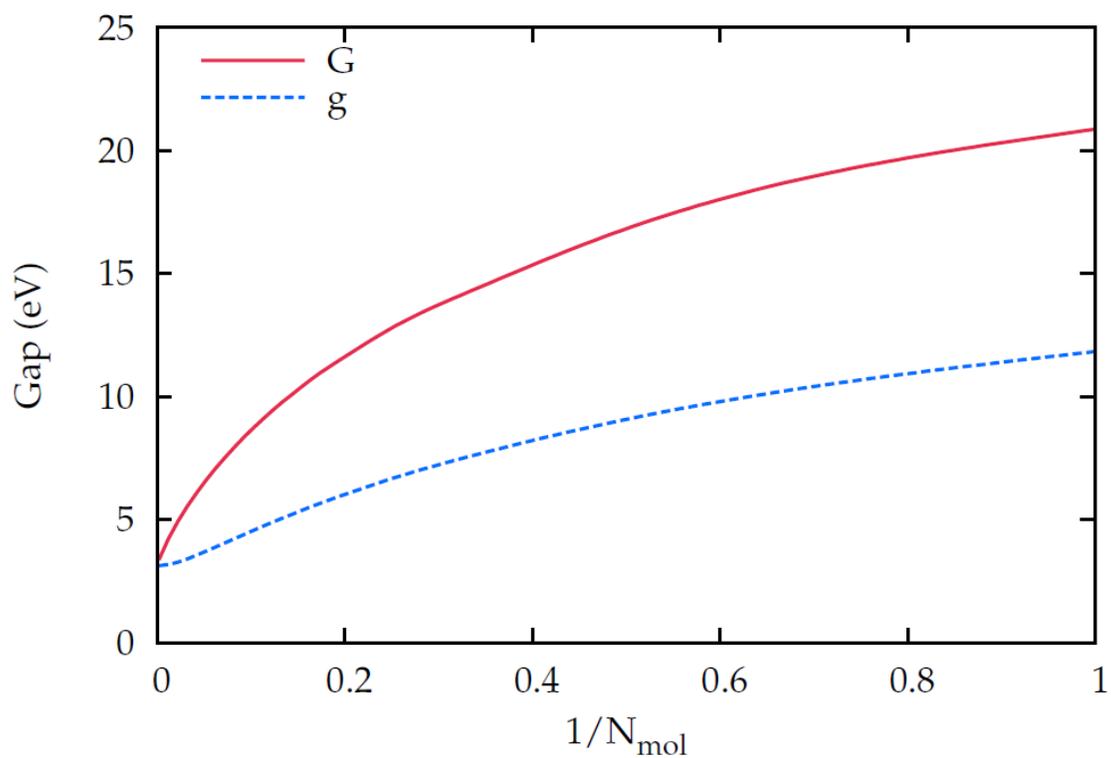

Fig. 1. The PBE GGA fundamental gap $G$ and band gap $g$ for a linear chain of $N_{mol}$ H$_2$ molecules. Note that $G$ converges to the limit $N_{mol} \to \infty$ much more slowly than $g$ does.

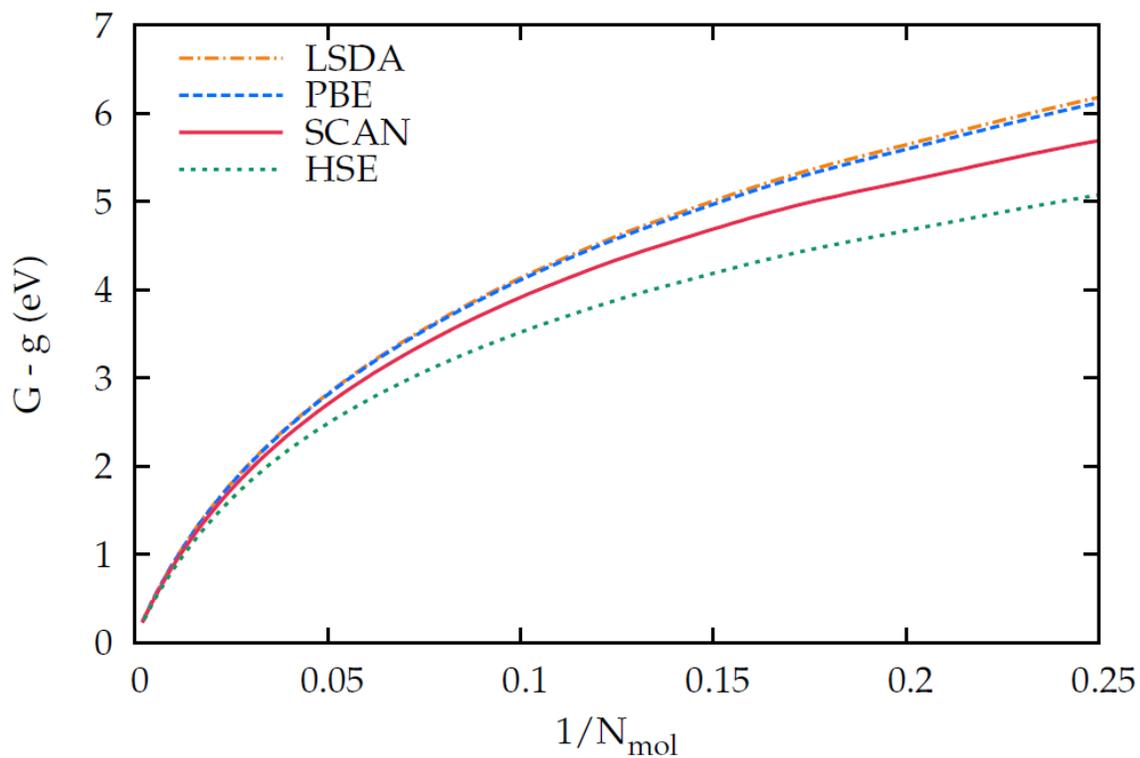

Fig. 2. Difference between the fundamental gap $G = I - A$ and the GKS band gap $g = \varepsilon^{LU} - \varepsilon^{HO}$ for a linear chain of $N_{mol}$ hydrogen molecules.

Table 1. Ionization energy $I$, electron affinity $A$, and fundamental gap $G = I - A$ of an infinite linear chain of H$_2$ molecules, evaluated by extrapolation from finite chains, and the band edges $\varepsilon^{HO}$, $\varepsilon^{LU}$ and band gap $g = \varepsilon^{LU} - \varepsilon^{HO}$, in the LSDA (2), PBE GGA (4), SCAN meta-GGA (5), and HSE06 range-separated hybrid (8) functionals. The extrapolated band energies agree closely with those from a periodic-boundary-condition calculation (shown). $(I + A)/2$, the energy difference from the gap center to the vacuum level (15), depends only weakly on the approximation.

| (eV) | $(I+A)/2$ | $I$ | $-\varepsilon^{HO}$ | $A$ | $-\varepsilon^{LU}$ | $G$ | $g$ |
|---|---|---|---|---|---|---|---|
| LSDA | 1.65 | 3.14 | 3.13 | 0.16 | 0.17 | 2.98 | 2.96 |
| PBE | 1.67 | 3.24 | 3.23 | 0.09 | 0.10 | 3.15 | 3.13 |
| SCAN | 1.68 | 3.33 | 3.31 | 0.01 | 0.02 | 3.32 | 3.29 |
| HSE06 | 1.82 | 3.92 | 3.91 | -0.29 | -0.28 | 4.21 | 4.18 |

Table 2. KS (PBE) and OEP/KS and GKS (PBE0) band gap $g$ and fundamental energy gaps $G$ of solid AlAs, calculated according to Eq. (1) with orbitals of the neutral $N$-electron system $G_{non\text{-}SCF}$, or with orbitals of separate self-consistent calculations of $N$-, $(N\text{-}1)$-, and $(N+1)$ - electron systems $G_{SCF}$, in eV, as described in the text. The experimental band gap (52) of AlAs is 2.23 eV.

|           | PBE   |             |           |           | PBE0  |             |           |
|-----------|-------|-------------|-----------|-----------|-------|-------------|-----------|
| Grid size | $g$   | $G_{non\text{-}SCF}$ | $G_{SCF}$ | $g_{OEP}$ | $g$   | $G_{non\text{-}SCF}$ | $G_{SCF}$ |
| 2x2x2     | 1.162 | 1.164       | 1.131     | 1.276     | 2.669 | 2.681       | 2.645     |
| 4x4x4     | 1.321 | 1.324       | 1.321     | 1.490     | 2.635 | 2.639       | 2.638     |
| 6x6x6     | 1.345 | 1.346       | 1.344     | 1.526     | 2.598 | 2.599       | 2.599     |
| 8x8x8     | 1.349 | 1.349       | 1.348     | 1.534     | 2.583 | 2.584       | 2.584     |
| 10x10x10  | 1.349 | 1.349       | 1.349     | 1.537     | 2.577 | 2.577       | 2.577     |
| 12x12x12  | 1.349 | 1.349       | 1.349     | 1.536     | 2.575 | 2.575       | 2.575     |

Table 3. KS(PBE) and OEP/KS and GKS(PBE0) band gap $g$ and fundamental energy gaps $G$ of solid Ar, calculated according to Eq. (1) with orbitals of the neutral $N$-electron system $G_{non\text{-}SCF}$, or with orbitals of separate self-consistent calculations of $N$-, ($N$-1), and ($N$+1)-electron systems $G_{SCF}$, in eV, as discussed in the text. The experimental band gap (53) of Ar is 14.20 eV. For a recent comparison of GKS band gaps for many solids from GGA hybrid functionals, including PBE0 and HSE, see Ref. 54.

| Grid size | PBE | | | | PBE0 | | |
| --- | --- | --- | --- | --- | --- | --- | --- |
| | $g$ | $G_{non\text{-}SCF}$ | $G_{SCF}$ | $g_{OEP}$ | $g$ | $G_{non\text{-}SCF}$ | $G_{SCF}$ |
| 1x1x1 | 7.621 | 9.130 | 8.482 | 7.901 | 12.079 | 11.944 | 11.311 |
| 2x2x2 | 8.640 | 8.793 | 8.658 | 8.831 | 10.947 | 11.065 | 10.948 |
| 3x3x3 | 8.688 | 8.735 | 8.694 | 8.923 | 11.091 | 11.108 | 11.073 |
| 4x4x4 | 8.691 | 8.711 | 8.699 | 8.938 | 11.120 | 11.123 | 11.108 |
| 5x5x5 | 8.692 | 8.702 | 8.693 | 8.942 | 11.121 | 11.126 | 11.119 |
| 6x6x6 | 8.692 | 8.697 | 8.693 | 8.944 | 11.122 | 11.126 | 11.122 |
| 7x7x7 | 8.692 | 8.695 | 8.692 | 8.945 | 11.123 | 11.126 | 11.123 |
| 8x8x8 | 8.692 | 8.694 | 8.692 | 8.945 | 11.123 | 11.125 | 11.124 |